\documentclass{article}
\usepackage{arxiv}
\usepackage{graphicx}
\usepackage{color}

\usepackage{paralist}
\usepackage[numbers]{natbib}
\usepackage[inline]{enumitem}
\usepackage{float}
\usepackage{url}

\usepackage{caption}
\makeatletter
\let\mcnewpage\newpage
\newcommand{\changenewpage}{%
  \renewcommand\newpage{%
    \if@firstcolumn
      \hrule width\linewidth height0pt
      \columnbreak
    \else
      \mcnewpage
    \fi
}}
\makeatother

\title{A Penalty-Aware Cloud Monitoring System based on Blockchains}
\vspace{6pt}

\author{
Christian Dienbauer\\
	Faculty of Computer Science\\
	University of Vienna, Austria\\
	\texttt{a1026759@univie.ac.at} \\
\AND
Benedikt Pittl\\
	Faculty of Computer Science\\
	University of Vienna, Austria\\
	\texttt{benedikt.pittl@univie.ac.at} \\
\AND
Erich Schikuta \\
	Faculty of Computer Science\\
	University of Vienna, Austria\\
	\texttt{erich.schikuta@univie.ac.at} \\
}

\renewcommand{\headeright}{}    
\renewcommand{\undertitle}{}    

\begin{document}

\date{}

\maketitle

\begin{abstract}
Today, traded cloud services are described by service level agreements that specify the obligations of providers such as availability or reliability. Violations of service level agreements lead to penalty payments. The recent development of prominent cloud platforms such as the re-design of Amazon's spot marketspace underpins a trend towards dynamic cloud markets where consumers migrate their services continuously to different marketspaces and providers to reach a cost-optimum. This leads to a heterogeneous IT infrastructure and consequently aggravates the monitoring of the delivered service quality. Hence, there is a need for a transparent penalty management system, which ensures that consumers automatically get penalty payments from providers in case of service violations. \newline
In the paper at hand, we present a cloud monitoring system that is able to execute penalty payments autonomously. In this regard, we apply smart contracts hosted on blockchains, which continuously monitor cloud services and trigger penalty payments to consumers in case of service violations. For justification and evaluation we implement our approach by the IBM Hyperledger Fabric framework and create a use case with Amazon's cloud services as well as Azures cloud services to illustrate the universal design of the presented mechanism.
\end{abstract}

\section{Introduction}        
Cloud providers such as Amazon and Azure sell their datacenter resources in form of services. For the description of services, so-called Service Level Agreements (SLAs) are used~\cite{PittlMS16iiwas}. They are specifications that define, inter alia, details about the offered service such as availability and reliability. Violations of SLAs can lead to penalty payments. Thus, consumers are interested in monitoring the service performance. Cloud providers such as  Amazons  offer  monitoring platforms like Amazon Cloudwatch that provide APIs to query the performance metrics of the service. Besides, several third-party cloud monitoring platforms are existing, such as dynatrace\footnote{https://www.dynatrace.com}, datadog\footnote{https://www.datadoghq.com/} or AppDynamics\footnote{https://www.appdynamics.com/}~.  Consumers use them to document service performance and to identify service anomalies which might result into a re-scheduling of workloads to other servers.
The scientific community introduced several penalty management approaches in the context of cloud computing ranging from penalty calculation~\cite{MaaroufQMH15} over penalty-aware resource management approaches to minimize costs~\cite{RahmanianDT17,Singh17} to an economical analysis of penalties~\cite{LuongWNWH17}. 
 The vision of a monitoring platform that also manages penalty payments is still unrivaled~\cite{NakashimaA17}. For the establishment of such a platform in industry not only machine-readable SLAs such as described in~\cite{NakashimaA17} are necessary, but also the transparent management of penalty payments has to be ensured. Hence, in the last years, neutral third parties were introduced that are responsible for confirming service violations and transferring penalty payments to consumers~\cite{MaaroufMMH18,ZhangLH17}. Even such neutral third party approaches have structural weaknesses as Robert Sams summarizes with three sins~\cite{mainelli2015sharing}: sin of commission, sin of deletion and sin of omission. Nowadays, the scientific community  reverts to neutral consensus finding approaches among untrusted participants which are technically realized by blockchain technology. For example, the authors of~\cite{UriarteNK18} introduced an approach for managing and modifying SLA terms with blockchains. However, penalty management for SLA violations was neglected. 
Based on them, reschedule workloads can be executed in order to avoid service interruptions. Furthermore, cloud monitoring is the precondition for penalty management: payments from the provider to the consumer in case of service violations.

The paper at hand focuses on the development of a blockchain-based, penalty-aware cloud monitoring platform. Thereby we apply so-called smart contracts, which are programs executed on the blockchain. A first step towards such systems was introduced in~\cite{neidhardtKN18}, where a pure  provider-centric smart contract was introduced. In contrast to that, the approach introduced in our paper treats a smart contract as a bilateral agreement between consumers and providers. Therefore, consumers and providers have to register their SLAs of traded cloud services in a smart contract, including corresponding commonly trusted cloud monitoring authorities. The smart contract uses them to monitor the performance of the cloud services and \emph{automatically} transfers penalties to the consumer in case of SLA violations. This ensures that consumers get  penalties immediately for service violations, while providers can profit from a clear and transparent monitoring approach. The technical feasibility of our approach is demonstrated by a smart contract deployed on the Hyperledger Fabric Blockchain\footnote{https://www.ibm.com/blockchain/hyperledger}~ that uses arbitrary monitoring services for detecting SLA violations and consequently executing penalty payments. Within this paper, we use cloud services as well as monitoring APIs both from Amazon and from Azure.

The remainder of the paper is structured as follows: First, we focus on foundations and related work. Then, we present the concept of the penalty-aware cloud monitoring platform, which is followed by an introduction of our implementation using the IBM Hyperledger Fabric Blockchain for Amazon EC2 and Azure. A discussion about the findings of the presented concept and future work is given followed by a summary and conclusion. We close the paper with an appendix which presents implementation details.


\section{Foundations and Related Work}
\label{sec:foundations}




This section is structured into two parts. The first part describes foundations of the blockchain technology and the second part presents related cloud monitoring approaches. 

With the rising popularity of the Bitcoin in mid-2017, the underlying blockchain technology gained attraction. This technology is not limited to cryptocurrencies and so organizations from various domains started to identify reliable business models enabled by the blockchain~\cite{beck2018beyond}. A typology of emerging blockchain applications with regards to the domains where they are applied is presented in~\cite{DBLP:conf/chi/ElsdenMBHSV18,zheng2018blockchain}. 
Due to the decentralized nature with no need for intermediaries, the blockchain technology can handle transactions in different markets and consequently reduces friction costs~\cite{crosby2016blockchain,beck2018beyond}. The Economist article~\emph{The Trust Machine} considers the blockchain technology as a solution for establishing  trust and its antecedents, as confidence, integrity, reliability, responsibility and predictability~\cite{beck2018beyond}. While sometimes blockchains are described as \textit{trust-less}~\cite{DBLP:conf/icis/AuingerR18}, other authors~\cite{DBLP:conf/hicss/LabazovaDS19} state that there is a shift of trust from current intermediaries to the technical implementation of blockchain systems. Limitations of the blockchain such as scalability, energy usage, and growing complexity are summarized in~\cite{welfare2019commercializing}. While data once stored within a blockchain is immutable, it does not mean that this data is always valid. Therefore, current research investigates in incentive-based blockchains to provide data quality~\cite{DBLP:conf/icis/AuingerR18}. When a blockchain needs to query data from the \emph{outside world}, it uses so-called \textit{oracles}. However, the capabilities of oracles are sometimes limited: For example, Ethereum\footnote{https://ethereum.org/}~ prohibits that smart contracts can query external sources, and so the data needs to be stored within a smart contract~\cite{bartoletti2017empirical} first. The smart contract with the injected data can then be used as oracle and queried by other smart contracts. Hyperledger Fabric, as a contrary example, uses general-purpose programming languages and allows that data can be retrieved from anywhere. Important hereby is that, no matter by whom this data is queried and regardless of time, the results need to be the same to ensure a deterministic behavior of the blockchain. 


Today, most of the monitoring platforms are cloud provider-specific limiting the monitoring of heterogeneous IT infrastructures, as found in smart city architectures and industry 4.0~\cite{DBLP:conf/closer/BayerMR19}. This enables data-driven applications that require reliable communication models to operate more efficiently~\cite{DBLP:conf/ewsn/JiangFP17}. Complex cloud service compositions are currently hard to monitor by existing monitoring platforms. In~\cite{DBLP:conf/IEEEcloud/NoorJMJSRD19} a generic framework to monitor the performance of services deployed on different providers is presented. To collect data from the services, the framework proposes a \emph{client-server approach}, where agents are deployed together with the service that should be monitored: Those agents can be accessed by a centralized monitoring platform to retrieve data of monitored services. A similar approach in the context of edge computing is presented in~\cite{DBLP:conf/closer/BayerMR19}. There, an architecture divided into management and worker layer is presented. The management layer provides functionalities to measure the activities of the edge nodes in the network to distribute workloads based on different algorithms. 

An approach for monitoring data streams in distributed systems with the focus on communication efficiency and data privacy can be found in~\cite{DBLP:conf/infocom/Sun0ZZ16}. The system collects information about different components and sets the granularity based on a given threshold by a centralized monitoring unit. Such centralized monitoring systems suffer from a single point of failure. Therefore, a distributed agent-based monitoring system to handle multi-tenant service-based systems is used~\cite{DBLP:conf/icws/YeHWY17}. A challenge of service monitoring is the system overhead of monitoring tools that the collection of data can be done in a resource-effective manner~\cite{DBLP:conf/icws/WangHYY17}.
The authors of~\cite{neidhardtKN18} introduce a concept of a blockchain based cloud monitoring system without a concrete implementation. Thereby, the provider creates  a smart contract that manages a list of consumers. In case of a service violation~\emph{service coins} are transferred to consumers. Motivated by the vision of billing consumers with smart contracts, the role of the consumer is neglected: it neither has to acknowledge the smart contract nor does it acknowledge the used monitoring services.

To foster the quality of services and the reliability of business processes that are based on these, Service Level Agreements (SLAs) between a service provider and consumer are defined~\cite{debe2019iot}. As SLAs are legally binding contracts, they have to be audited by an instance that is trusted by all participants~\cite{weishaupl2006gset}. While conventional SLA trust models are centrally configured, deployed and maintained, new approaches to define SLAs using smart contracts can be found in~\cite{alzubaidi2019blockchain, scheid2018automatic, nakashima2017automation,uriarte2018towards, debe2019iot, wonjiga2019blockchain, ULHAQ2011435}.
A conceptual blockchain-based framework for SLA management is presented in~\cite{alzubaidi2019blockchain}. It gives a general overview of this topic and addresses the challenge of trust between all participants arguing that no single party should have control over the SLA lifecycle. In traditional SLA approaches service consumers need to detect abnormalities and check SLA violations, which leads to an extensive manual process including personnel and capital resources~\cite{alzubaidi2019blockchain, nakashima2017automation}. An approach for automated compensation of SLAs based on smart contracts can be found in~\cite{scheid2018automatic}. It compares its approach to existing solutions with regards to human-computer interaction. 
 
A formal description of a SLA lifecycle based on a smart contract is presented in~\cite{uriarte2018towards} that includes 5 steps: 
\begin{enumerate*}
  \item Discovery of service and Negotiation of SLAs,
  \item Deployment of the SLA using smart contracts,
  \item Monitoring of the service,
  \item Billing and Penalty Enforcement, and
  \item Termination of the smart contract.
\end{enumerate*}
A system architecture in the context of fog computing can be found in~\cite{debe2019iot} that provides IoT client devices with the means to explore the best-suited fog nodes via smart contracts by considering the reputation and credibility based on SLA requirements.

Current approaches try to address today's requirements of heterogeneous and decentralized infrastructures by agent-based systems with either a centralized unit or a wholly distributed approach. They identify the challenges of data privacy and communication efficiency. These approaches are usually isolated systems as the data is subject to a single entity with lack of transparency. Thus, for organizations it is hard to get reliable information about the current and past state of their IT infrastructure~\cite{vossen2012cloud}. Consequently, consumers are forced to implement systems to gather information This results in additional costs and the burden of proofing the lack of quality~\cite{alzubaidi2019blockchain, nakashima2017automation}. 

While approaches exist to define Service Level Agreements using smart contracts without the need to rely on a single party and automate the way to identify SLA violations, they are either conceptual or lack the implementation of a penalty mechanism. Approaches mentioned in~\cite{alzubaidi2019blockchain,scheid2018automatic} use Ethereum as the underlying blockchain technology but miss the part on how to inject service metrics into smart contracts to trigger events while preserving a deterministic behavior of the blockchain.
Therefore, we focused on the implementation of a monitoring system for a service provider where SLA can be defined. Compared to existing approaches we implemented a mechanism to define SLAs and trigger penalties in case of their violation. 

Thus, for our research endeavor we pursue the following requirements:

\begin{enumerate}
    \item It is mandatory that cloud services from multiple service provider can be monitored.
    \item The monitoring tool needs to be distributed with no single point of failure.
    \item Data within the system will be treated confidentially.
    \item There is no need to trust in a single participant and transparency of the system.
    \item Implementation uses effective communication models for scalability. 
\end{enumerate}

If trust in the system exists, such systems have the potential to automatically settle dispute based on terms participants agreed on earlier.

\begin{figure}[htbp]
    \centering
    \includegraphics[width=0.6\linewidth]{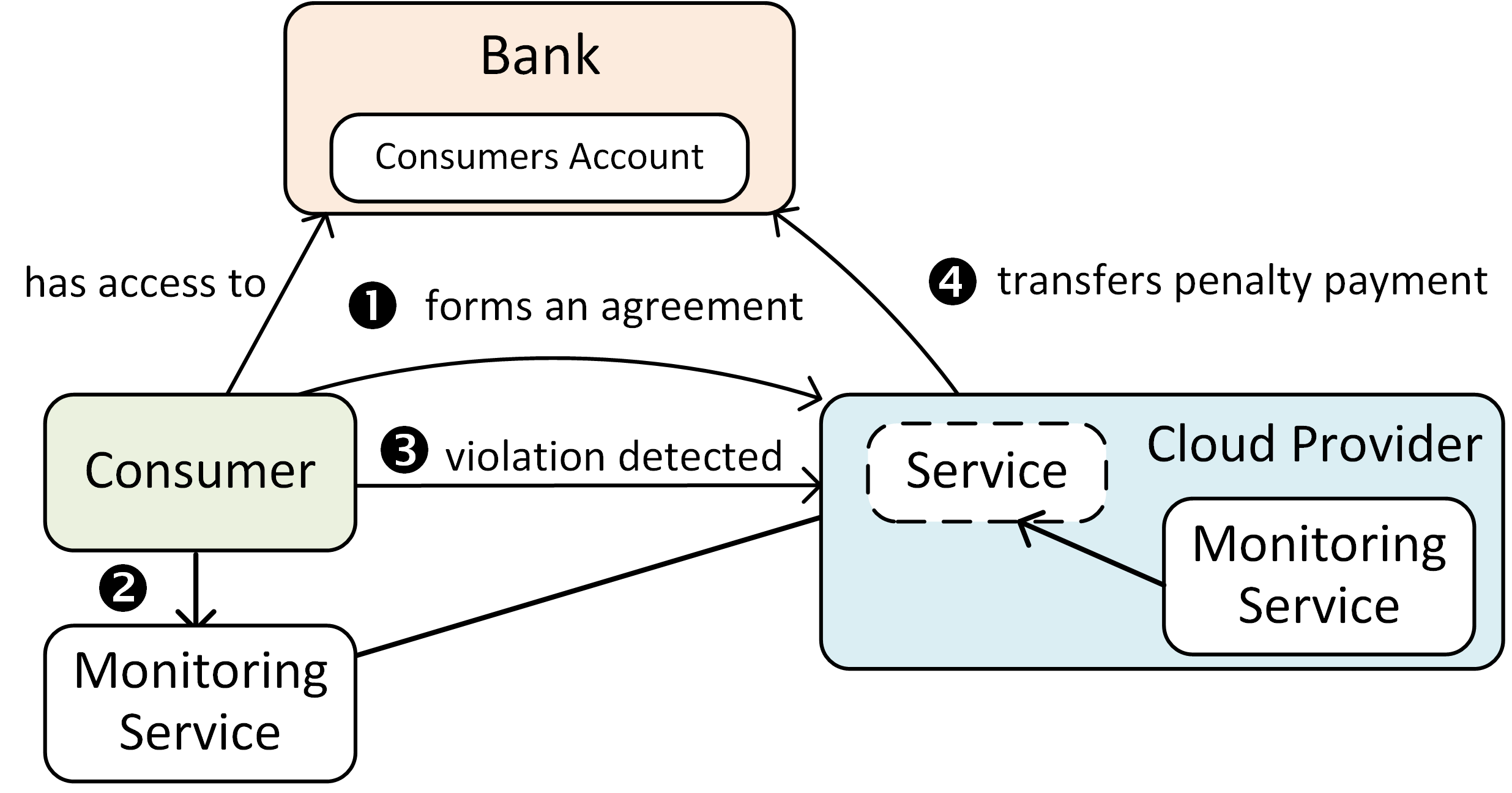}
    \caption{Traditional cloud monitoring scenario}
    \label{fig:trad}
\end{figure}

\section{Cloud Monitoring with Smart Contract}
\label{sec:core}

Today, consumers purchase cloud resources from few providers. More dynamic cloud marketspaces are evolving where providers and consumers apply multi-round negotiation processes for trading cloud resources~\cite{ChichinVK17,PittlHMS19,PittlMS16,pittl2015negotiation}. Initiatives such as the Deutsche Börse Cloud Exchange, where consumers were able to choose between appropriate providers for required cloud services, underpin the trend towards dynamic cloud markets. Thereby, consumers purchase cloud services on-demand from preferable providers. On such marketspaces, providers can increase the utilization of their cloud datacenter while consumers profit from lower prices and customized cloud services~\cite{PittlMS16}. However, consumers will have multiple cloud providers as business partners which they have to maintain. This leads to challenges concerning cloud monitoring and penalty-management. Hence, a transparent and autonomous system for penalty management is necessary so that consumers can utilize cloud services from multiple, probably untrusted, cloud providers without putting remarkable effort into service monitoring and penalty management.

A traditional cloud monitoring scenario is presented in Figure~\ref{fig:trad}. First, the consumer purchases a service from a provider (1) then both, consumer and provider use services to monitor the performance of traded service (2). As soon as the consumer observes a service violation, it claims a penalty payment from the provider (3). Finally, the provider transfers the penalty payments to the consumers' bank (4). The traditional monitoring scenario comes with two drawbacks: First, the identification of a service violation can lead to contradictory results as both, the consumer and the provider could use separate monitoring services. Second, the scenario requires a trusted third party such as the bank which is responsible for transferring penalty payments.

To tackle these issue,s the paper introduces a cloud monitoring system based on blockchains. It uses smart contracts that are responsible for processing data from monitoring services and executing penalty payments in case of identified cloud service violations. The following two components are essential for such a smart contract:
\begin{itemize}
    \item \textbf{Monitoring Service.}
    
    A smart contract cannot monitor the cloud service directly. Therefore, it has to make use of a monitoring service that returns performance metrics. Such services are offered by cloud providers such as Amazon\footnote{https://aws.amazon.com/de/cloudwatch/}, but also by independent monitoring service providers such as datadoghq\footnote{https://www.datadoghq.com}~. Both, the consumer and the provider have to agree on the selected monitoring service. Even multiple monitoring services could be used in parallel, whereby the smart contract has to decide which value is used in the case of contradictory results. In the rest of the paper, the monitoring services are explicitly referred to as~\emph{monitoring services} to distinguish them from the traded cloud service, which is termed~\emph{service}.
    
    \item \textbf{Wallet.}
    
    The smart contract is a neutral entity that is trusted by both, consumer and provider. To guarantee transparent and fast penalty payments, the smart contract needs a wallet. Instead of directly transferring the service fee  to the provider, the consumer has to pay it by transferring tokens\footnote{Tokens can be money but also other trade-able entities.}~ to the smart contract. Those tokens are stored in the wallet of the smart contract. In the case of service violations, tokens are transferred back to the consumer. The remaining tokens are transferred to the provider after the contract period expired.
\end{itemize}

\begin{figure}[htbp]
    \centering
    \includegraphics[width=0.8\linewidth]{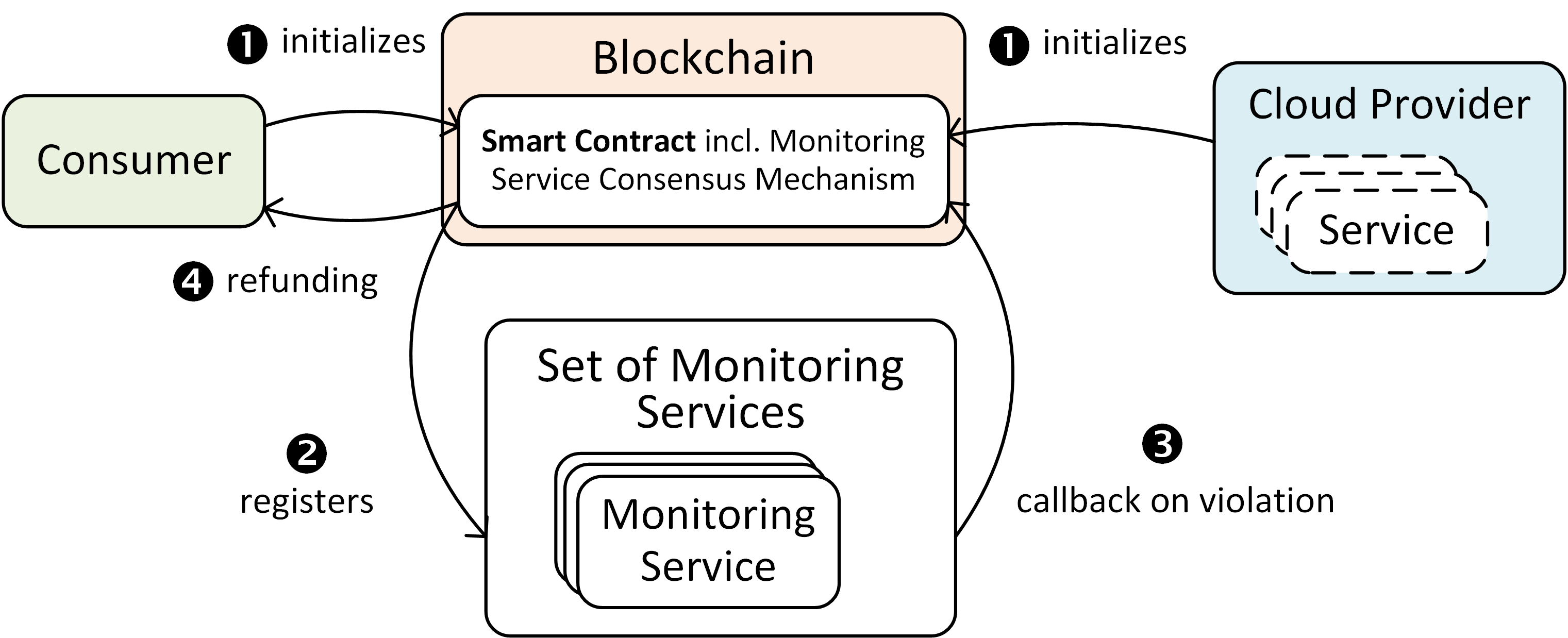}
    \caption{Blockchain-based cloud monitoring scenario}
    \label{fig:overview}
\end{figure}

Figure~\ref{fig:overview} introduces an exemplary scenario in which a~\emph{cloud provider} hosts the service used by the~\emph{consumer}. The numbers in the following enumeration refer to the numbers in the figure.

\begin{enumerate}
    \item A consumer decided to use a cloud service from a provider for a certain period of time. For the autonomous penalty management, the consumer and the provider create a smart contract. The provider adds endpoints of the used monitoring services to it. They are used by the smart contract for monitoring the service performance. The consumer agrees to use the service by transferring the service fee in form of tokens to the smart contract. They are stored in the smart contract's wallet. 
    
    \item The consumer uses the service. The service is monitored by the  monitoring services defined in the smart contract. 
    
    \item Service violations detected by the monitoring service are accessible for the smart contract. If multiple monitoring services are used, the smart contract has to resolve contradictory results to determine if a service violation occurred. This conflict resolution is part of the smart contract, so both the consumer and the provider are aware of it.
    
    \item If the smart contract detects a service violation, it automatically transfers penalty payments from its wallet to the consumer. After the contract period between consumers and providers expired, the remaining tokens in the wallet of the smart contract are transferred to the provider. If no service violations occur, all tokens representing the service fee are transferred to the provider.
\end{enumerate}

\begin{figure}[htbp]
    \centering
    \includegraphics[width=0.6\linewidth]{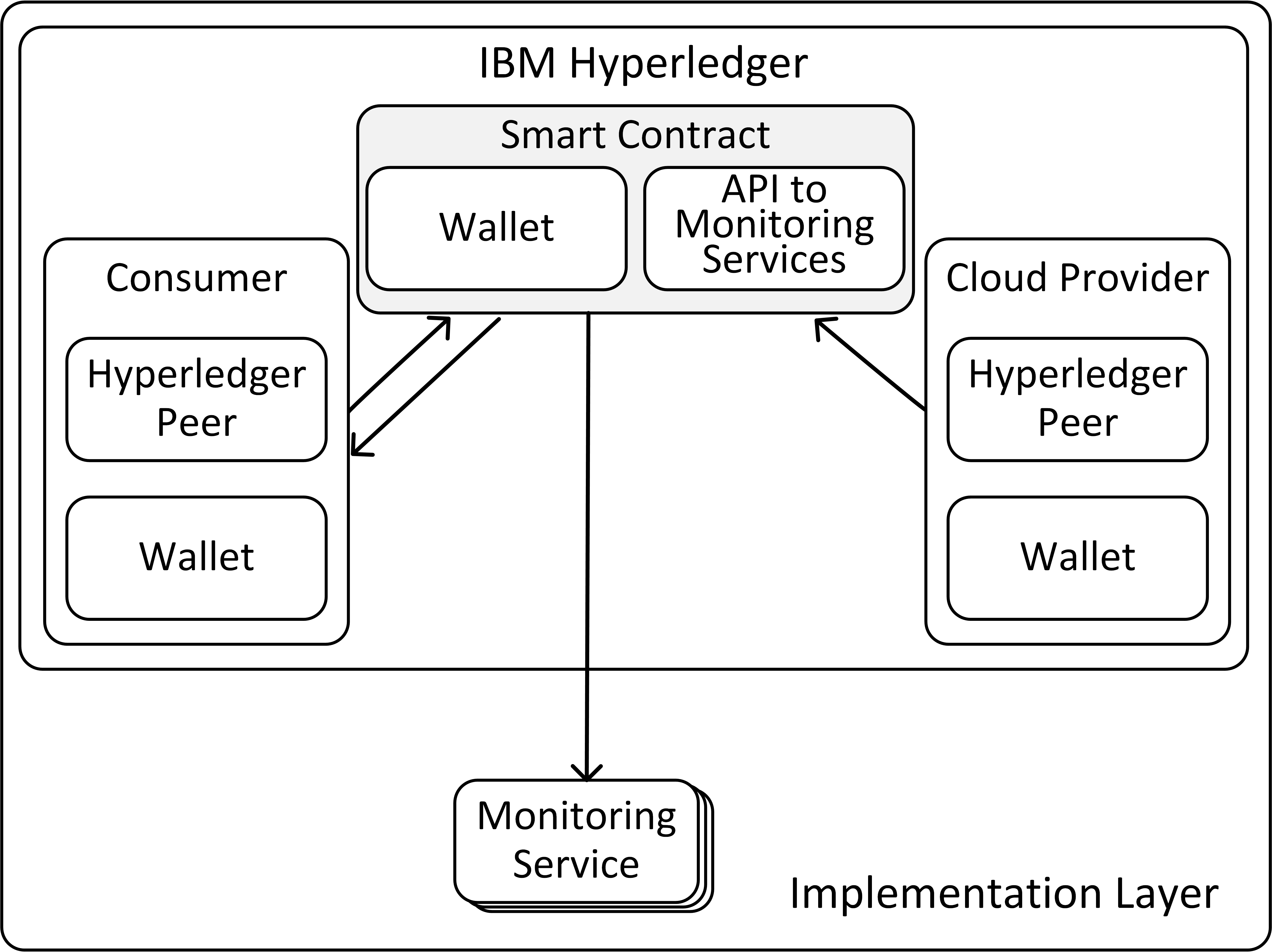}
    \caption{Implementation layer of the monitoring scenario}
    \label{fig:impllayer}
\end{figure}

The implementation layer of the conceptual design of Figure~\ref{fig:overview} can be realized according to Figure~\ref{fig:impllayer}.

The introduced approach has several benefits over approaches where non-blockchain based software solutions are employed for penalty management. 
\begin{inparaenum}[(i)]
\item First of all, smart contracts are deployed and executed on the blockchain and consequently on a neutral platform while typical software solutions have to be hosted on a server. There, the availability, as well as the execution, are not guaranteed.
\item Further, smart contracts are participants of the blockchains, which means that they can send and receive tokens such as money.   For the realization of non-blockchain based software, a trusted third party such as a bank would be necessary.
\item Trusted third parties charge fees and require time for executing payments. On the blockchain, these fees for trusted third parties are not relevant. The consumer gets penalty payments automatically and immediately on detected service violations.
\item Smart contracts are immutable, which means that the code cannot be modified after it is deployed and so the approach is appropriate for untrusted consumers and providers.
\end{inparaenum}

The consumers have to transfer the service fee upfront to the smart contract. This implies that the consumer has to pre-pay the complete service fee while the provider gets the remaining service fee at the end of the contract period. Especially for high service fees or long contract periods, this might cause financial embarrassment. In such cases, the consumer and the provider could agree to transfer only parts of the payments to the smart contract. 

\subsection{Forming an Agreement}
\noindent  We assume that the deployment of the Smart Contract takes place during the subscription process for a service. Customers  lookup the desired service via the product catalog of a service provider and choose available subscription options. In our example of subscribing to a virtual machine, customers can select the number of CPUs, the available RAM, or the available network bandwidth as well as metrics for the availability. The service properties and selected options are then used to shape the characteristics of the smart contract to enforce SLA violations. Important hereby is that metrics stated in an SLA, need to be requestable by a smart contract using the monitoring tools of the service provider as an oracle.

\begin{figure}[htbp]
    \centering
    \includegraphics[width=0.7\textwidth]{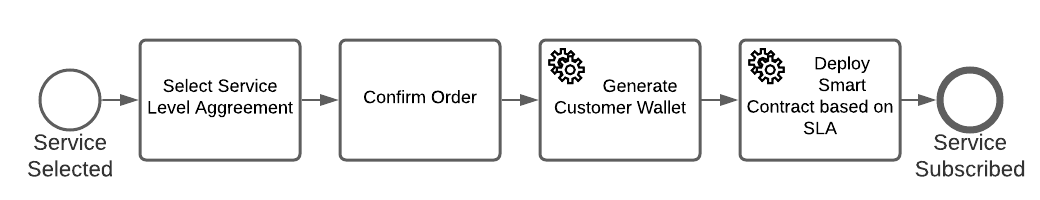}
    \caption{Defining SLAs in a Smart Contract}
    \label{oracle_figure1}
\end{figure}

After the customer accepted the Service Level Agreements, a wallet holding the cryptographic key-pair will be generated for the customer. The smart contract with the defined SLAs and information about the wallet addresses of the provider and the client will be deployed to the blockchain network (see Figure~\ref{oracle_figure1}). All blockchain-related steps are performed automatically in the background without any interaction needed by the customer. At the end of the subscription process, the customer  gets a summary of the deployed smart contract and information on how to access the wallet. 
Executing the contract verifies that the service operates as stated in the SLAs. If it fails to deliver the characteristics as defined during the subscription process, a penalty payment is transferred to the wallet of the customer.
This corresponds to step two, three, and four of the SLA lifecycle described in~\cite{uriarte2018towards}.

\subsection{Retrieving Data Off-Chain}
\noindent  Sometimes blockchain-based systems need to obtain information outside the boundaries of the blockchain network. This is necessary, if volatile information, e.g. data from the stock market or metrics from a web service, are used as a basis for decisions.  While data once stored in the blockchain is immutable and therefore considered as trustworthy, using data from the~\emph{outer-world} is a critical process to preserve trust and transparency. For example, an approach could mitigate the strength of a blockchain, where the client application sends information as a payload to a smart contract. No other participant of the network could be able to verify if the information a user committed is correct.

This challenge is addressed by so-called~\emph{oracles}. An oracle is a data source queried by smart contracts influencing their action. Depending on the underlying blockchain technology an oracle can either be:

\begin{enumerate}
    \item Smart contract itself, or
    \item Trusted third party service
\end{enumerate}

Important for oracles is their deterministic behavior. Thus, regardless of time, the participants of the network and who is querying the smart contract, the oracle will always deliver the same result.

In the case of Ethereum, an oracle is a smart contract, as Solidity, the language used to implement smart contracts, has no concept to query external services. So off-chain data needs to put into a smart contract first, which can afterward be queried by other smart contracts. The correctness of the data put into the smart contract makes it reproduce- and traceable. The smart contract that acts as oracle can implement some logic, that the input of multiple parties will be aggregated. For example, if an escrow service is implemented, an oracle will consider the input of the buyer, the seller, and a trusted third party. If the seller and buyer do not agree, the input of a trusted third party will be required to either send the funds to the seller or give them back to the buyer.

\begin{figure}[htbp]
    \centering
    \includegraphics[width=0.5\textwidth]{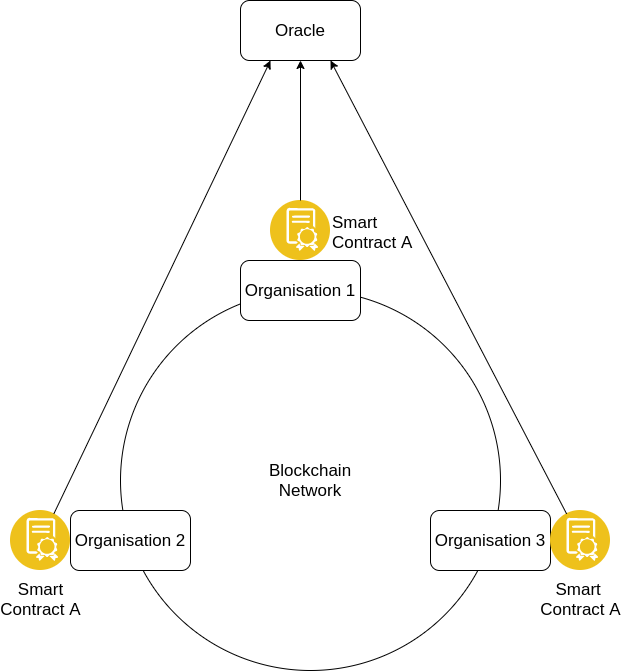}
    \caption{Participants validate data}
    \label{oracle_figure}
\end{figure}

Other blockchain technologies, e.g. Hyperledger, use general-purpose programming languages like JavaScript, Python, or Go to request services. In this case, an oracle can be a trusted third party service that is queried in runtime. The logic  is implemented in a smart contract on which the participants of a network previously agreed. Thereby, it is important to preserve the deterministic behavior of the blockchain again. 
In figure~\ref{oracle_figure} the concept is shown, where multiple organizations execute the same smart contract to verify that the data stored in the blockchain is indeed the data from the trusted third party.

\section{Implementation}
\label{sec:implementation}

\begin{figure}[htbp]
    \centering
    \includegraphics[width=0.6\textwidth]{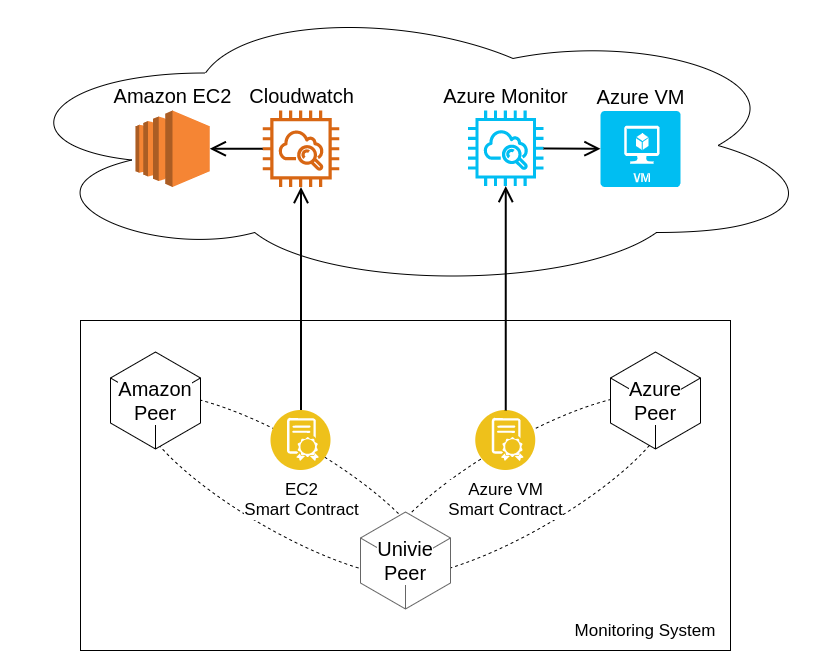}
    \caption{Scenario of the cloud monitoring system}
    \label{fig:scenario}
\end{figure}

\begin{figure}[htbp]
    \centering
    \includegraphics[width=0.8\textwidth]{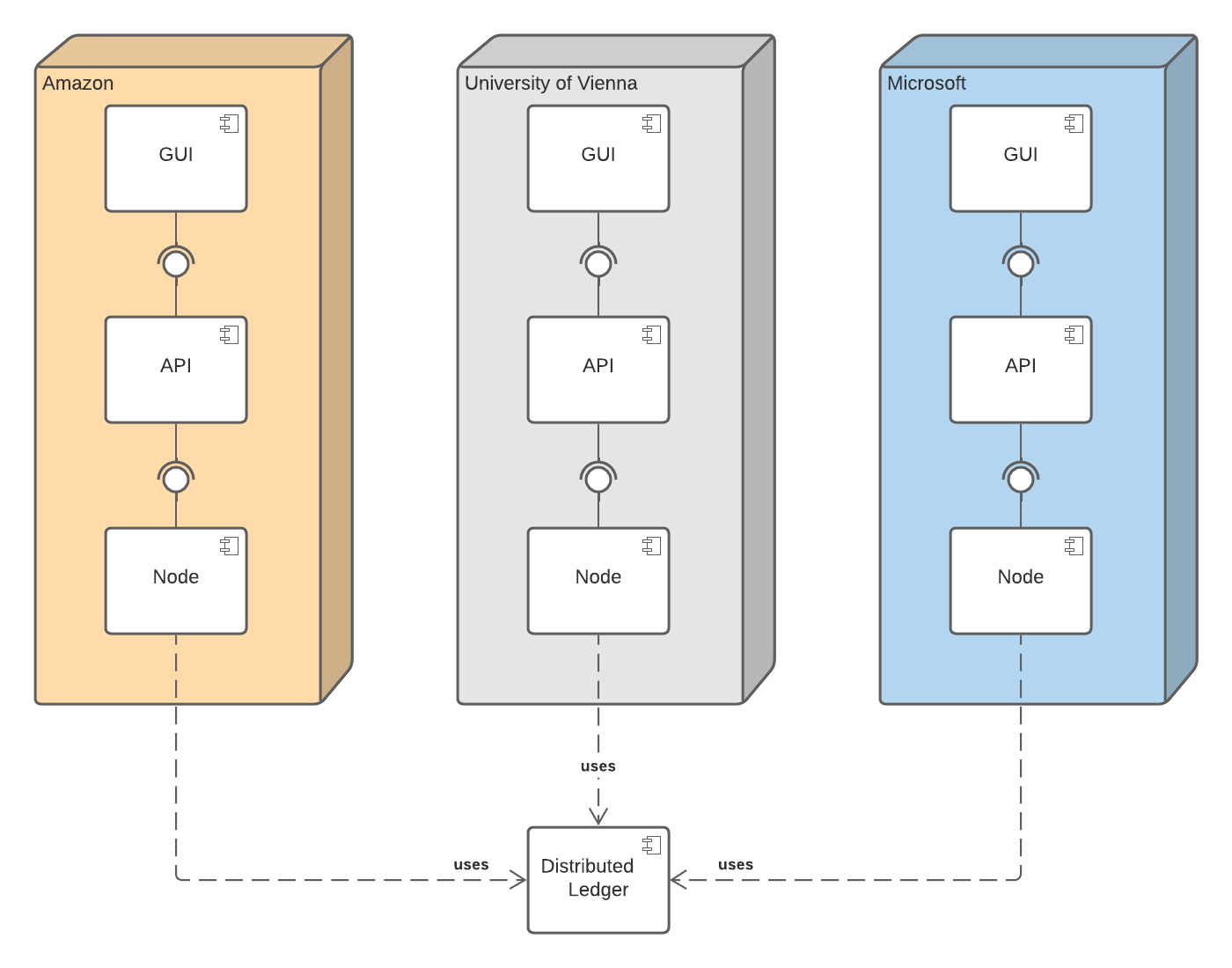}
    \caption{Components of the cloud monitoring system}
    \label{fig:components}
\end{figure}

This section introduces the implementation of the scenario shown in Figure~\ref{fig:scenario}. It distinguishes between three stakeholders: two cloud service providers and one consumer. In the scenario, one service provider is Amazon, the other service provider is Microsoft Azure. The consumer uses a virtual machine, which represents the traded cloud service from both providers. For each cloud service, a smart contract is instantiated. It has access to monitoring services hosted by the corresponding provider respectively. Those monitoring services act as oracles for the smart contract and allow for a deterministic behavior of the states in the blockchain. The components of this scenario are depicted in Figure~\ref{fig:components}.

Different blockchain technologies can be used for the implementation of the scenario. For the selection of appropriate blockchain technology implementing the envisioned penalty-aware cloud monitoring platform, we follow the blockchain-selection guideline~\cite{wust2018you}. As the participants' identities will be established before purchasing a service and data of services is not publicly verified, we decided to use a private permissioned blockchain.  A private permissioned blockchain will guarantee that service data will only be accessible for the users specified in an endorsement policy: the service provider and the consumer. Transaction fee for processing data is not relevant for private blockchains. 
For the presented use-case we used the Hyperledger Fabric framework\footnote{https://www.hyperledger.org/use/fabric}~, which supports the usage of a general-purpose programming language for writing smart contracts, chaincode as they call them, to implement any conceivable logic.

The blockchain network for this experiment can be created with the use of some tools provided by Hyperledger. These tools can be used via the command line interface to create fully functional blockchain network. 

\begin{figure}
    \centering
    \includegraphics[width=0.9\linewidth]{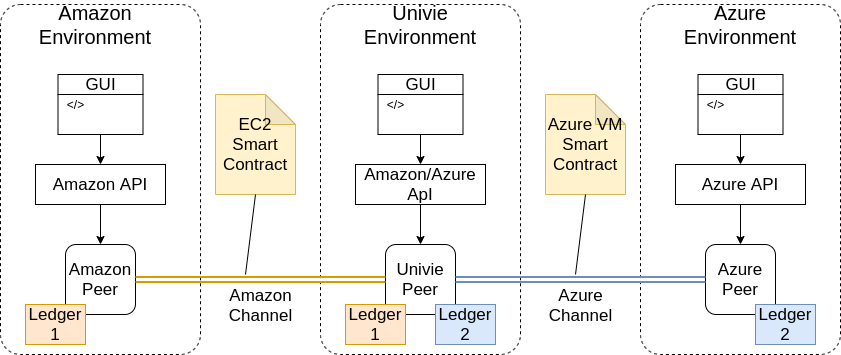}
    \caption{Implementation of the concept}
    \label{fig:architecture}
\end{figure}

Figure~\ref{fig:architecture} shows the technical architecture of the implemented scenario. It consists of the smart contracts, monitoring service integration code and the core environment. 

Peer Nodes, or short peers, are a fundamental part of a blockchain network and inherit a cryptographic identity given by a certificate authority. With this identity, peers can be authenticated in the network and, with regards to the endorsement policy, join certain channels. A channel is a sub-net of the blockchain network and enables private communication between peers. Members of the channel share a distributed ledger to store confidential transactions. Smart contracts are deployed on channels and can be executed by its peers to interact with the ledger. In our example, the service consumer shares a separate ledger with each provider.
Each participant deploys a graphical user interface and an API via a docker-compose file in its environment. Communication with the participants of the network is only done via smart contracts. 

\begin{itemize}
    \item \textbf{Smart Contract.}  
    
    For the implementation of the smart contracts we decided to use JavaScript, as Hyperledger Fabric officially supports an SDK for~\emph{Node.js}. This SDK enables to interact with the blockchain network by using an API to submit transactions to, and query information from the ledger. 
If a consumer consumes multiple services from a provider, one smart contract is sufficient. It can handle multiple services of the same type by linking the measurements to a single instance via a unique service key. In the code snippet below an excerpt of a smart contract is shown for retrieving measurements. The credentials are stored in the ledger and will be used for querying the Amazon EC2 Oracle.

\begin{verbatim}
async updateService(ctx, serviceKey, startTime, endTime) {
    const serviceAsBytes = await ctx.stub.getState(serviceKey); 
    if (!serviceAsBytes || serviceAsBytes.length === 0) {
        throw new Error(`${serviceKey} does not exist`);
    }
    let service = JSON.parse(serviceAsBytes.toString());
    ...
    // query the oracle
    service.data = await this.getMetricsDataEC2(
        service.accessKeyId,
        service.secretAccessKey,
        service.region,
        service.dimensionName,
        service.dimensionValue,
        startTime,
        endTime
    );
    // calculate availability
    if (service.cpuUtilization >= 0) {
        service.timesAvailable += 1;
        console.log("Connection good");
    } else {
        service.timesNotAvailable += 1;
        console.log("No connection");
    }
    ...
    // give credit if availability is to low
    if (currentAvailability < service.promisedAvailability) {
        service.credit += PENALTY_AMOUNT;
    }
    await ctx.stub.putState(
        service.name,
        Buffer.from(JSON.stringify(service)));
\end{verbatim}

    \item \textbf{Monitoring Service Integration.} 
    
    Smart contracts require access to the monitoring services. In blockchain-environments, accessing such external services is technically challenging: The~\emph{Node.js} SDK provided by Amazon is used to retrieve metrics of the service from their monitoring tool Cloudwatch. 
    When a service is registered to the monitoring system, an unique service key with all the necessary information to retrieve metrics from Cloudwatch is stored within the ledger. This enables retrieving information from Cloudwatch and storing the information into the ledger running purely inside a smart contract by just passing the service key for identification. To query information from the ledger we implemented some methods to retrieve all registered service keys and to retrieve all historical service states about a specific service from the ledger. 
    
    An update function is used to retrieve the information about the service metrics of the latest time interval and store them into the ledger. But it would also be possible to upgrade Cloudwatch capabilities to retrieve metrics in a granularity of down to a minute. 
    During registration of a service in our system, the availability promised by the service provider is passed as a parameter and stored together with the credentials and information of the service in the ledger. As the update function queries the information about the service from Cloudwatch, we see from the response if a service is available or not. Within the smart contract we mark this requested interval as either available or unavailable and compare it to the total run-time of the service. If a service's downtime exceeds the promised availability of a service provider, the customer will receive a certain amount of credit for this service. This credit can then be used as an offset during settlement.
    As smart contracts are not self-executable programs, the update function needs to be requested from outside of the network. To prevent a user from requesting the service's status for the same time interval twice, smart contracts check at each request of the update function if the time of the last state in the ledger will not overlap with the time interval in the request.
    
    The same logic is used for the implementation of the smart contract for Azure VM services. However, instead of using an SDK to retrieve information from the Azure monitoring tool, we decided to use their provided REST API.
\newline    
    
    \item \textbf{Core Environment.} 
    
    To provide users of the system a convenient way to register services and explore their current and past state together with actual availability and possible credits, we created a graphical user interface with React. As React is based on JavaScript and the Hyplerdeger also provides a JavaScript SDK for application to interact with smart contracts deployed in the network, our first attempt was to use this SDK within the React application. Unfortunately this was not possible as the version we used of React is based on the ES6 JavaScript specification, whereby the SDK for Fabric 1.4 follows the specification of ES5. This mismatch of versions of the specifications lead to import errors and crashes of the React application. 
    Thus, we developed a REST API which uses the SDK to interact with the blockchain network and provides endpoints that can be requested by the React application. This API has as further advantage that it can be used by a cron job to automatically query the latest states of a registered service.
\end{itemize}

A reference to the complete script to reproduce the created setup can be found on GitHub\footnote{https://github.com/dieni/blockchain-based-cloud-monitoring}~. The main components of the script are:

\begin{enumerate}\label{script}
    \item Generate x.509 certificates,
    \item Configuration of the channels,
    \item Creation of the peer nodes,
    \item Joining the peers to the channels, and
    \item Instantiating the smart contracts in the channels
\end{enumerate}

\begin{figure}[htbp]
    \centering
    \includegraphics[width=0.9\linewidth]{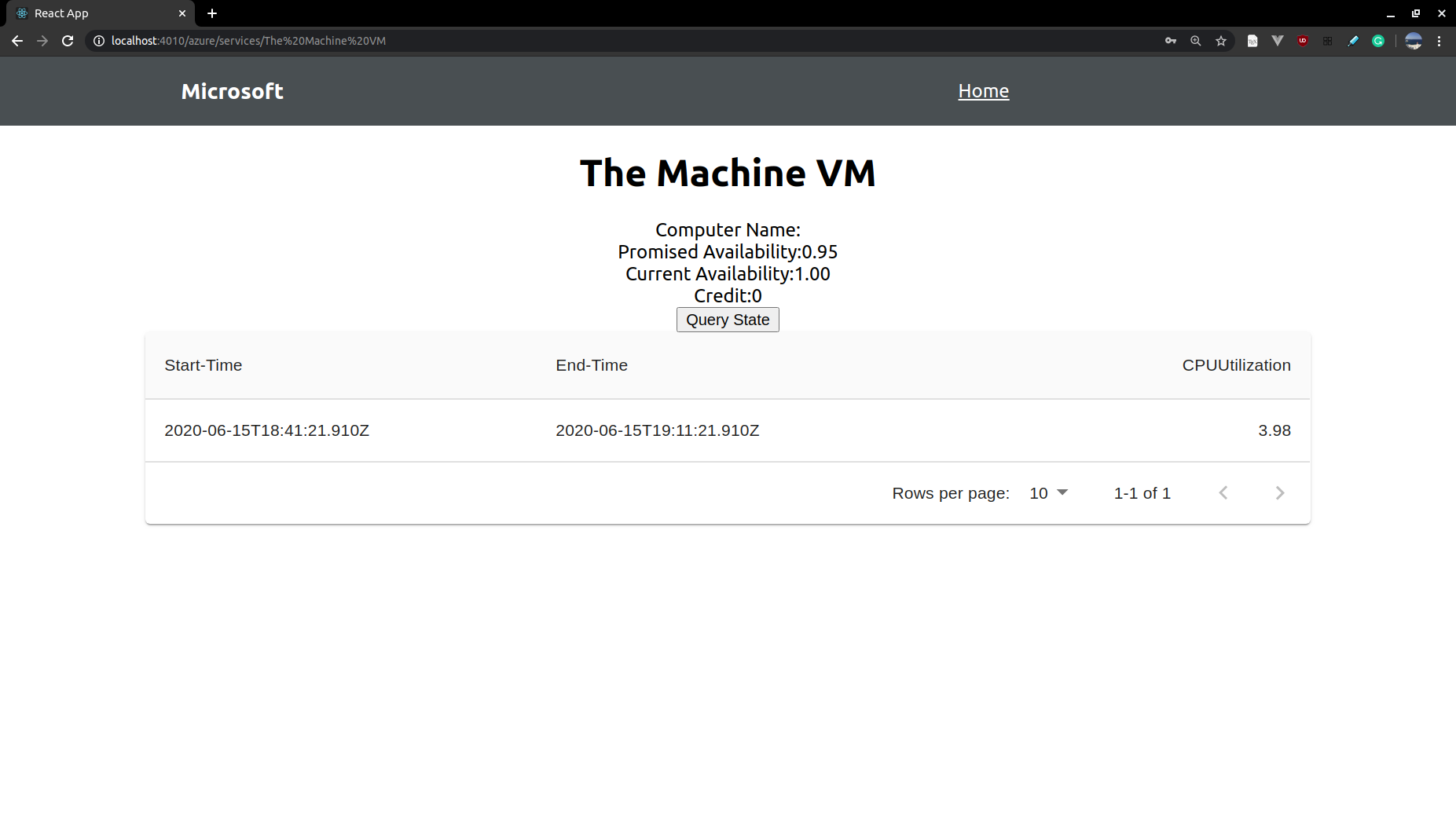}
    \caption{Service details page that show the current status of the monitored service - the same information is retrieved by the smart contract}
    \label{fig:details_page}
\end{figure}

The current implementation of the monitoring service is capable of retrieving information about two service types of two different service providers. To prove the feasibility of integrating service level agreements we started with the implementation of the \textit{promised availability} that is passed to the system during registration process. Though, as this logic is deployed via smart contracts, the system can be extended in a modular manner to support additional service types as well, as a variety of service metrics that are provided by the oracles. All this information can then be used to define SLAs via the monitoring system. 
If an SLA agreement is violated, the current approach will trigger a penalty by increasing the credit of variables within the ledger that is shared between the service provider and the consumer. To change the value of this variable, a predefined logic within the smart contract is used that considers all the past states of service as well as the promised availability passed during registration of the service in the system.
As smart contracts are not self-executable programs, the process of retrieving new states of a service needs to be started by the user. Currently, this is done via a button in the user interface but can be automated using a cron job. For this purpose, we created a RESt API as it was necessary to implement the JavaScript SDK to interact with the smart contracts. As first attempt we used the SDK within the React application, which led to importing errors, as the JavaScript standard used by React (ES6) did not match the standard used by the SDK (ES5). The code listing 
shows the endpoint of the API that is used by the frontend to query the smart contract to retrieve new states of service. The user interface is deployed locally without any direct connection to the internet. All the information that is shown is retrieved from the blockchain network by using smart contracts. The code snippet shows how new services are registered to the system. Our scenario uses services from Amazon and Microsoft and can choose between two registration forms. This form contains information for authentication to the oracles as well as information about the SLA. If the entered information is correct, the credentials will be stored in a smart contract. On the bottom of the form, a list of all registered services of a provider can be seen. When clicking on the services, the details page of the service~\ref{fig:details_page} opens: Here a user can see information about the retrieved measurements of the service. By clicking on the button, new measurements can be retrieved. If the service is more often unreachable than promised, the customer will receive a credit.

\begin{verbatim}

    app.put('/aws/services/:serviceKey', async (req, res) => {
      console.log(req.body)
      let serviceKey = JSON.parse(req.params.serviceKey)
    
      const gateway = new Gateway()
      
      try {
        console.log('CONNECT TO NETWORK')
        await gateway.connect(connectionProfile, connectionOptions)
        const network = await gateway.getNetwork('channel-aws')
        const contract = network.getContract('ec2contract')
    
        const endTime = new Date()
        const startTime = new Date(endTime.getTime() - 30 * 60000)
    
        await contract.submitTransaction(
          'updateService',
          serviceKey,
          startTime.toISOString(),
          endTime.toISOString()
        )
    
        console.log('SERVICE UPDATED')
      } finally {
        // Disconnect from the gateway
        console.log('Disconnect from Fabric gateway.')
        gateway.disconnect()
      }
    })

\end{verbatim}

\begin{figure}[htbp]
    \centering
    \includegraphics[width=0.9\textwidth]{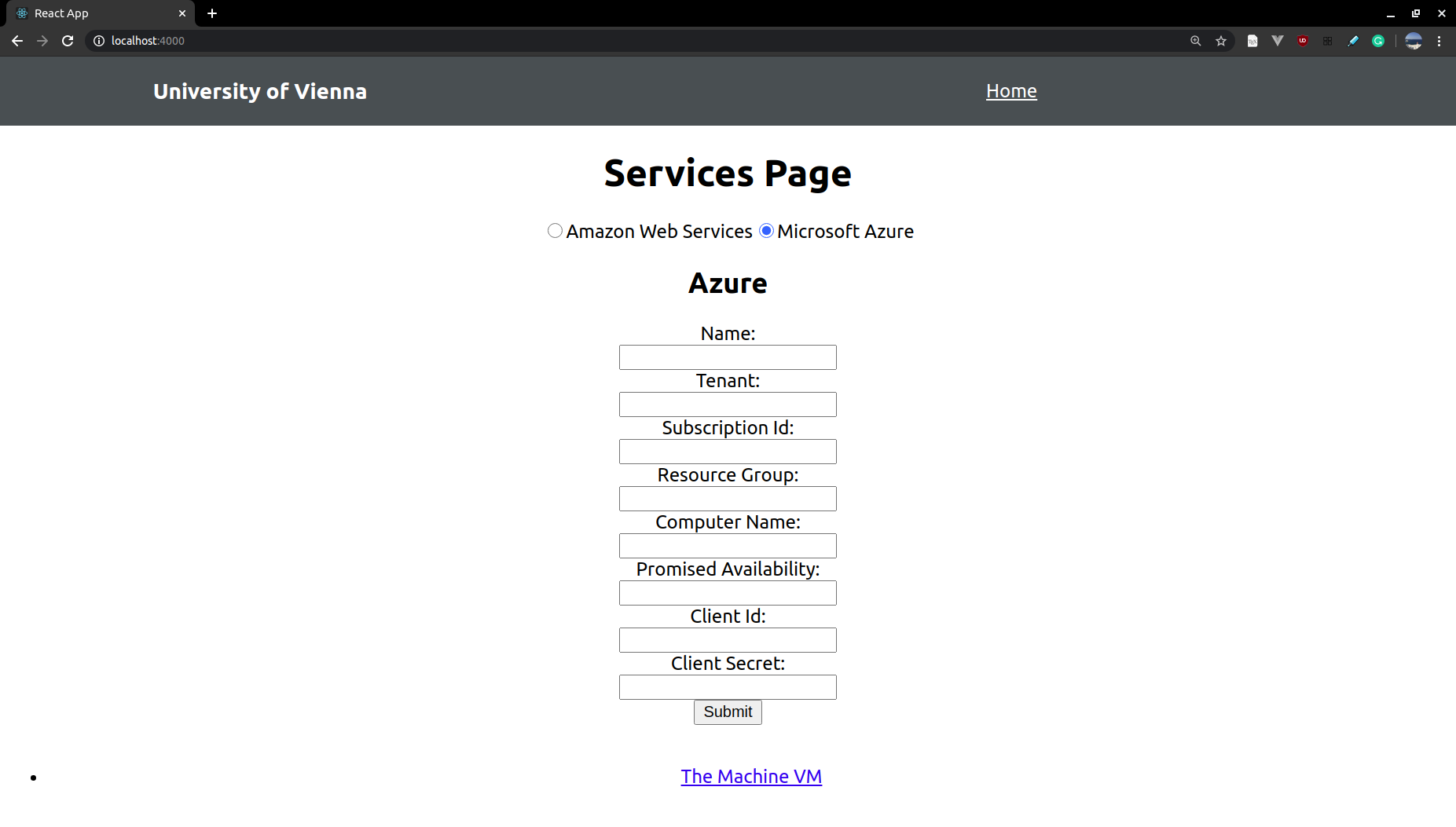}
    \caption{Registration page of a cloud service that should be monitored by the smart contract deployed on the Hyperledger Fabric Network}
    \label{fig:reg_page}
\end{figure}

The certificates are necessary to authenticate the organizations in the network. In a production environment organizations need to use their own and keep then confidential. Each channel has its own ledger and with a configuration of an endorsement policy is defined who is allowed to join the channel. As channels are established between each service provider and the customer(s), it is ensured that no service provider has access to the information about the services of the other provider. The instantiated smart contracts of in the channels contains the logic to retrieve data from the oracles and define the interaction with the ledger. 

In the appendix we document the system setup as well as the available services that we implemented.

\section{Findings and Future Work}
\label{sec:findings}
\noindent  This paper is part of a research project aiming on using blockchain technology for monitoring services of different service providers. Thereby, we aim at a decentralized architecture to enable penalty payments if service level agreements are violated. This concept is realized by the presented implementation, where for each service type a smart contract is deployed in a self hosted Hyplerledger Fabric network and shared between service provider and consumer via private channels. Even more, the design of the implemented approach will work for any other provider as long as an API is available that returns the required service performance metrics. A pre-condition of the introduced concept is, that consumers and providers participate in a blockchain network such as the presented Hyperledger Fabric Network. 

The following taxative list sums up the findings of the presented work and gives impetus for future work:
\begin{itemize}
    \item \textbf{Blockchain as a Service}
    
    The network was created using the CLI tools provided by Hyperledger and was deployed locally running the components within docker containers. Hyperledger provides  the containers via docker hub for all necessary components to run the network. Though, the configuration of the network can be quite difficult as documentation about different settings and operations is quite rare. A recommended approach for a future project is to use Blockchain as a service. Amazon~\footnote{https://aws.amazon.com/marketplace/pp/B0797GK9YY}~ as well as Microsoft~\footnote{https://azuremarketplace.microsoft.com/en-us/marketplace/apps/microsoft-azure-blockchain.azure-blockchain-hyperledger-fabric-aks-based}~ provide managed blockchains services to create Hyperledger Fabric networks as well as Ethereum. They claim that this can be done with minimal time for configuration. Another platform to be considered for using Hyperledger Fabric as a service is from IBM~\footnote{https://www.ibm.com/blockchain}~. 
    
    \item \textbf{Smart Contracts.}
    
    With the use of a general-purpose programming language, the development of smart contracts is nothing out of the ordinary. Hyperledger Fabric supports a chaincode SDK for JavaScript that enables a convenient interaction with the ledger using a so-called \textit{transaction context}. Other languages are supported too. Packages can be used without any restrictions. 
    One thing to keep in mind when developing smart contracts is, that each operation needs to follow a deterministic behavior. This is necessary that nodes of the network reach consensus and the states in the ledger are reproducible.
    
    \item \textbf{Oracles.}
    
    For the deterministic behavior of the system, we used the monitoring tools of the service provider as our oracles. Amazon as well as Microsoft provides a sophisticated API to query current and past states of services based on different metrics. For AWS we used a JavaScript SDK and for Azure a Rest API. Both methods work like expected and can be used to query information in granularity of one minute. Currently, we query only one metric to determine the availability of service. This can be extended to a variety of metrics with an convenient way to select which should be monitored. 
    The execution of a request will be done by a single node and the results distributed through the network. To prevent the owner of a node making changes to the request, the oracle might sign the reply with its private key so that other nodes can verify the validity of the data.
    
    
    \item \textbf{Multiple service types.}
    
    We decided to deploy one smart contract between a service provider and consumer for each service type. This enables a consumer to monitor multiple services from the same provider using a single smart contract. It has to be investigated how to enable the monitoring of multiple services from different types between a service provider and consumer. One approach could use one smart contract for each additional service type that has to be monitored. However, This results in a very static design and a large number of smart contracts when supporting several service providers  each delivering a variety of services. 
    A generic approach of registering different service types is worthwhile to be investigated, to limit the number of smart contracts and to have a more flexible way of on-boarding new service providers with there services. 
    
    \item \textbf{Service Level Agreements.}
    
    For this project, we used a single metric to determine the availability of the service and compared it to the promised availability. With each request to an oracle, the state of the service was compared with all past events. As smart contracts are not self-executable, we had to care that requests are not made twice for the same time interval. We solved this by checking the last time interval stored in the ledger before retrieving new measurements from the oracle. 
    For more sophisticated SLA's we will need to consider more metrics and let the users define the appropriate levels for those. The best case would be to combine this task with a generic approach to registering services.  
    
    \item \textbf{Penalty Payments.}
    
    As already said, if there is a violation of the SLA's, a predefined amount is credited to the consumer. This is currently done via a variable stored in the ledger that can only be modified based on the retrieved metrics from the oracles. While the credit is defined when creating the invoice, we aim for an independent process of transferring funds between the users. One approach is to define a penalty token that will be transferred between the users, if SLA's are violated. Therefore we need to replace the logic of increasing the variable with the mechanism to trigger the transfer. The tokens need to be transferred between wallets dedicated to the users to give them the possibility to have them at one's disposal later.
    
    \item \textbf{Tokens.}
    
    Usually, in blockchain networks tokens Bitcoins are transferred instead of money. They represent compensation of money and can be transferred and traced in a tamper-resistant manner. Especially the ability of smart contracts to transfer tokens~\emph{in the program code} enables the implementation of the introduced monitoring system. On the contrary, if consumers and provider do not execute any other transactions in that blockchain network, they have to exchange them for global currencies such as Dollars or Euros. Especially for industrial use-cases this comes with currency risks as the exchange rate is volatile. State-supported Blockchain networks could help to tackle that issue.
    
\end{itemize}

\section{Summary and Conclusion}
\label{sec:conclusion}
\noindent  The trend towards dynamic cloud markets, where consumers purchase cloud services from various providers in an ad-hoc manner, leads to heterogeneous IT-infrastructures. Consumers have to ensure that the service performance conforms to the published service specification as violations might lead to penalty payments. With the increasing number of potentially unknown service providers the penalty management becomes a relevant issue for consumers, which raises the need of an autonomous penalty-aware cloud monitoring system that ensures transparent penalty management in case of service violations.

In the presented paper, we applied blockchain technology for the realization of such a monitoring system: Traded cloud service are registered in a smart contract, which continuously monitors the performance of the traded cloud service. As soon as a service violation is identified, the smart contract transfers penalty payments to the consumer. Consumers profit from an autonomous transfer of penalty payments while providers profit from a transparent and reasonable assessment of services. Due to the management of the penalty payments in the smart contracts, correct payoffs are ensured for both, the provider and the consumer.

To prove the technical feasibility of the approach, a penalty-aware cloud monitoring system was implemented using the IBM Hyperledger Fabric framework. The implemented scenario, monitoring the availability of the virtual machines from Amazon and Azure, illustrates the generic applicability of our introduced approach. For immediate adoption in industry a unification of service performance metrics is necessary as well as a predictable exchange rate for tokens to global currencies. 

\bibliographystyle{IEEEtranN}
\bibliography{allCites}


\section*{Appendix}

\subsection*{Setup of the System}
The implementation of the system can be found on GitHub\footnote{https://github.com/dieni/blockchain-based-cloud-monitoring}~.

The creation of the blockchain network is done with some command-line tools provided by Hyperledger. All the commands to spin up the network including peer nodes for the users are composed in \textit{sms.sh} within the network folder. This script file takes one of the three options: up, down, and generate.

\begin{enumerate}

    \item Install prerequisites for Hyperledger Fabric as described in its documentation
    
    \item Navigate into the network folder and execute the command \textit{./sms.sh up}\\
    This will execute multiple operations:
    \begin{itemize}
        \item If they do not already exist, this will generate certificates for each of the organizations and stores them into the \textit{crypto-config} folder
        
        \item Based on the endorsement policy stated in \textit{configtx.yaml} the artifacts for the channel creation will be generated and stored in the folder \textit{channel-artifacts}. Note: When generating the artifacts, the folder needs to exist already.
        
        \item Starts docker service containers described in the \textit{docker-compose.yaml} file which are an ordering service, peer nodes for each organization and another peer node which acts as a helper for further configuration of the network
        
        \item Within the cli peer the \textit{utils.sh} the script will be executed which makes use of the crypto files of the different organizations and performs following operations:
        
        \begin{itemize}
            \item Based on the previously generated channel artifacts two channels will be created and registered to the ordering service: \textit{channel-aws} and \textit{channel-azure}
            
            \item The AWS peer together with the peer of University are joined to the channel-aws
            
            \item Analog to the previous step the Azure peer together with the peer of the University are joined to the channel-azure
            
            \item The smart contract for the EC2 services gets installed on the Univie and AWS peer and instantiated on the channel-aws
            
            \item The smart contract for the Azure VMs gets installed on the Univie and Azure peer and instantiated on the channel-azure
        \end{itemize}
        
        \item When you see in the terminal following output, the creation of the network was successful:\\
        '========= SMS Up! ==========='

\end{itemize}
        
    \item Generate wallets for each of the organizations\\
    This is necessary for the SDK to interact with the ledger

    \begin{enumerate}
        \item Go into the \textit{crypto-config} folder and copy the name of the .pem file of the organization you would like to interact with the network.\\
       \path{/network/crypto-config/peerOrganization/ <organization>/users/<user>}
       
       \path{/msp/keystore/<...9a08817cd53f6_sk>}\\

       Beware that the different users of an organization have different permissions. Those were stated in the endorsement policy on the creation process of the channel artifacts. For the reading and writing \textit{User1}.
        
            
        \item Within the folder of the organization open the file: \\
        \path{organization/{organization}/application/addToWallet.js}\\
        
        \begin{itemize}
            \item Past the name of the .pem file and make sure that the path to the .pem file matches the actual location.
            \item Define the path to the MSP certificate of the identity form the \textit{crypto-config}
            \item Define a label for the identity and save the file
            
        \end{itemize}
        
        \item Execute the \textit{addToWallet.js}\\
            This will create a wallet within the folder of the organization which can be used for an application to authenticate to the peer and manage transactions
        
        \item Repeat this procedure for all three organizations
        
    \end{enumerate}
        
    \item Execute the \textit{docker-compose} file with the organization folder\\
    This will create the container for the User Interface and the API of each organization.
    The user interfaces for the organizations can be found under:
    
        \begin{itemize}
            \item localhost:4000 (univie)
            \item localhost:4010 (azure)
            \item localhost:4020 (aws)
        \end{itemize}
    
\end{enumerate}

\subsection*{API of the System}
The implementation is located in \textit{/organizations/{org}/server} and uses the Nodejs library 'express' for building a REST API. Depending on the organization this interface consists either of the routes for the AWS service, for the Azure service, or both in the case of the University of Vienna. The APIs can be accessed under the following addresses:

\begin{itemize}
    \item localhost:4001 (University of Vienna)
    \item localhost:4011 (Azure)
    \item localhost:4021 (Amazon)
\end{itemize}

The endpoints for EC2 instances of the API are as following:

\begin{table}[H]
\centering
\begin{tabular}{|p{3.5cm}|p{7.5cm}|} 
\hline
\multicolumn{2}{|l|}{\textbf{/aws/services}}                                                                                                                         \\ 
\hline
Method               & GET                                                                                                                                 \\ 
\hline
\multicolumn{2}{|p{11cm}|}{Returns the keys of all EC2 instances registered in the ledger of the AWS channel. It uses the function \textit{getallKeys} of the AWS EC2 contract. This operation does not change the state of the ledger.}    \\
\hline
Request   & -\\
\hline
Response   & HEADER:\\
& Content-Type: application/json\\
& \\
& BODY:\\
& List of service keys\\ 
\hline
\end{tabular}
\end{table}

\begin{table}[H]
\centering
\begin{tabular}{|p{3.5cm}|p{7.5cm}|} 
\hline
\multicolumn{2}{|l|}{\textbf{/aws/services}}                                                                                                                         \\ 
\hline
Method               & POST                                                                                                                                 \\ 
\hline
\multicolumn{2}{|p{11cm}|}{Register a EC2 instance in the ledger of the AWS channel. Uses the function \textit{createService} of the AWS EC2 contract. This operation changes the state of the ledger.}    \\ 
\hline
Request   & HEADER:\\
& Content-Type: application/json \\
& \\
& BODY:\\
& \{ \\
& \hspace{2ex}name: $<string>$,\\
& \hspace{2ex}dimensionName: $<string>$,\\
& \hspace{2ex}dimensionValue: $<string>$,\\
& \hspace{2ex}region: $<string>$,\\
& \hspace{2ex}accessKeyId: $<string>$,\\
& \hspace{2ex}secretAccessKey: $<string>$,\\
& \hspace{2ex}promisedAvailability: $<string>$\\
& \{ \\
\hline
Response   & - \\
\hline
\end{tabular}
\end{table}

\begin{table}[H]
\centering
\begin{tabular}{|p{3.5cm}|p{7.5cm}|} 
\hline
\multicolumn{2}{|l|}{\textbf{/aws/services/$<serviceKey>$}}                                                                                                                         \\ 
\hline
Method               & GET                                                                                                                                 \\ 
\hline
\multicolumn{2}{|p{11cm}|}{By passing the service key in the url the ledger will be queried for all measurements of a service. It uses the function \textit{getStateHistory} of the AWS EC2 contract. This operation does not change the state of the ledger.}    \\ 
\hline
Request   & -\\
\hline
Response   & HEADER:\\
& Content-Type: application/json\\
& \\
& BODY:\\
& List of measurements
\\ 
\hline
\end{tabular}
\end{table}

\begin{table}[H]
\centering
\begin{tabular}{|p{3.5cm}|p{7.5cm}|} 
\hline
\multicolumn{2}{|l|}{\textbf{/aws/services/$<serviceKey>$}}                                                                                                                         \\ 
\hline
Method               & PUT                                                                                                                                 \\ 
\hline
\multicolumn{2}{|p{11cm}|}{By passing the service key in the url information about a service will be retrieved from the service provider and written to the ledger. It uses the function \textit{updateService} of the AWS EC2 contract. This operation changes the state of the ledger.}    \\ 
\hline
Request   & -\\
\hline
Response   & -\\ 
\hline
\end{tabular}
\end{table}

\vspace*{20px}

The endpoints for Azure VM of the API are as following:

\begin{table}[H]
\centering
\begin{tabular}{|p{3.5cm}|p{7.5cm}|} 
\hline
\multicolumn{2}{|l|}{\textbf{/azure/services}}                                                                                                                         \\ 
\hline
Method               & GET                                                                                                                                 \\ 
\hline
\multicolumn{2}{|p{11cm}|}{Returns the keys of all Azure VM instances registered in the ledger of the Azure channel. It uses the function \textit{getallKeys} of the Azure VM contract. This operation does not change the state of the ledger.}    \\
\hline
Request   & -\\
\hline
Response   & HEADER:\\
& Content-Type: application/json\\
& \\
& BODY:\\
& List of service keys\\ 
\hline
\end{tabular}
\end{table}

\begin{table}[H]
\centering
\begin{tabular}{|p{3.5cm}|p{7.5cm}|} 
\hline
\multicolumn{2}{|l|}{\textbf{/azure/services}}                                                                                                                         \\ 
\hline
Method               & POST                                                                                                                                 \\ 
\hline
\multicolumn{2}{|p{11cm}|}{Register a Azure VM instance in the ledger of the Azure channel. Uses the function \textit{createService} of the Azure VM contract. This operation changes the state of the ledger.}    \\ 
\hline
Request   & HEADER:\\
& Content-Type: application/json \\
& \\
& BODY:\\
& \{ \\
& \hspace{2ex}name: $<string>$,\\
& \hspace{2ex}tenant: $<string>$,\\
& \hspace{2ex}clientId: $<string>$,\\
& \hspace{2ex}clientSecret: $<string>$,\\
& \hspace{2ex}subscriptionId: $<string>$,\\
& \hspace{2ex}resourceGroup: $<string>$,\\
& \hspace{2ex}computerName: $<string>$,\\
& \hspace{2ex}promisedAvailability: $<string>$\\
& \{ \\
\hline
Response   & - \\
\hline
\end{tabular}
\end{table}

\begin{table}[H]
\centering
\begin{tabular}{|p{3.5cm}|p{7.5cm}|} 
\hline
\multicolumn{2}{|l|}{\textbf{/azure/services/$<serviceKey>$}}                                                                                                                         \\ 
\hline
Method               & GET                                                                                                                                 \\ 
\hline
\multicolumn{2}{|p{11cm}|}{By passing the service key in the url the ledger will be queried for all measurements of a service. It uses the function \textit{getStateHistory} of the Azure VM contract. This operation does not change the state of the ledger.}    \\ 
\hline
Request   & -\\
\hline
Response   & HEADER:\\
& Content-Type: application/json\\
& \\
& BODY:\\
& List of measurements
\\ 
\hline
\end{tabular}
\end{table}

\begin{table}[H]
\centering
\begin{tabular}{|p{3.5cm}|p{7.5cm}|} 
\hline
\multicolumn{2}{|l|}{\textbf{/azure/services/$<serviceKey>$}}                                                                                                                         \\ 
\hline
Method               & PUT                                                                                                                                 \\ 
\hline
\multicolumn{2}{|p{11cm}|}{By passing the service key in the url information about a service will be retrieved from the service provider and written to the ledger. It uses the function \textit{updateService} of the Azure VM contract. This operation changes the state of the ledger.}    \\ 
\hline
Request   & -\\
\hline
Response   & -\\ 
\hline
\end{tabular}
\end{table}


\end{document}